\documentclass[sigconf,twocolumn, screen]{acmart}
\AtBeginDocument{%
  \providecommand\BibTeX{{%
    \normalfont B\kern-0.5em{\scshape i\kern-0.25em b}\kern-0.8em\TeX}}}

\setcopyright{acmlicensed}
\copyrightyear{2024}
\acmYear{2024}
\acmDOI{XXXXXXX.XXXXXXX}

\acmConference[--]{--}{--}{--}
\acmISBN{--}

\begin{document}

\title{Pack my weights and run! Minimizing overheads for in-memory computing accelerators}


\author{Pouya Houshmand}
\affiliation{%
  \institution{KU Leuven}
  \city{Leuven}
  \country{Belgium}}
\email{pouya.houshmand@kuleuven.be}

\author{Marian Verhelst}
\affiliation{%
  \institution{KU Leuven}
  \city{Leuven}
  \country{Belgium}}
\email{marian.verhelst@kuleuven.be}


\begin{abstract}
In-memory computing hardware accelerators allow more than 10x improvements in peak efficiency and performance for matrix-vector multiplications (MVM) compared to conventional digital designs. For this, they have gained great interest for the acceleration of neural network workloads. 
Nevertheless, these potential gains are only achieved when the utilization of the computational resources is maximized and the overhead from loading operands in the memory array minimized. To this aim, this paper proposes a novel mapping algorithm for the weights in the IMC macros
, based on efficient packing of the weights of network layers in the available memory. The algorithm realizes 1) minimization of weight loading times while at the same time 2) maximally exploiting the parallelism of the IMC computational fabric. A set of case studies are carried out to show achievable trade-offs for the MLPerf Tiny benchmark \cite{mlperftiny} on IMC architectures, with potential $10-100\times$ EDP improvements.
\end{abstract}


\maketitle

\section{Introduction}
\label{sec:intro}

In recent years there has been a massive surge of interest for the development of hardware accelerators for neural networks for running models on the edge, where compute and memory resources are limited. Due to the large amount of compute and memory required for such workloads, Von Neumann processors are not enough anymore to satisfy the power and throughput requirements of most modern AI workloads at the edge -- mostly dominated by matrix-vector multiplications (MVMs). 

\begin{figure}[!t]
    \centering
    \includegraphics[width=0.9\linewidth]{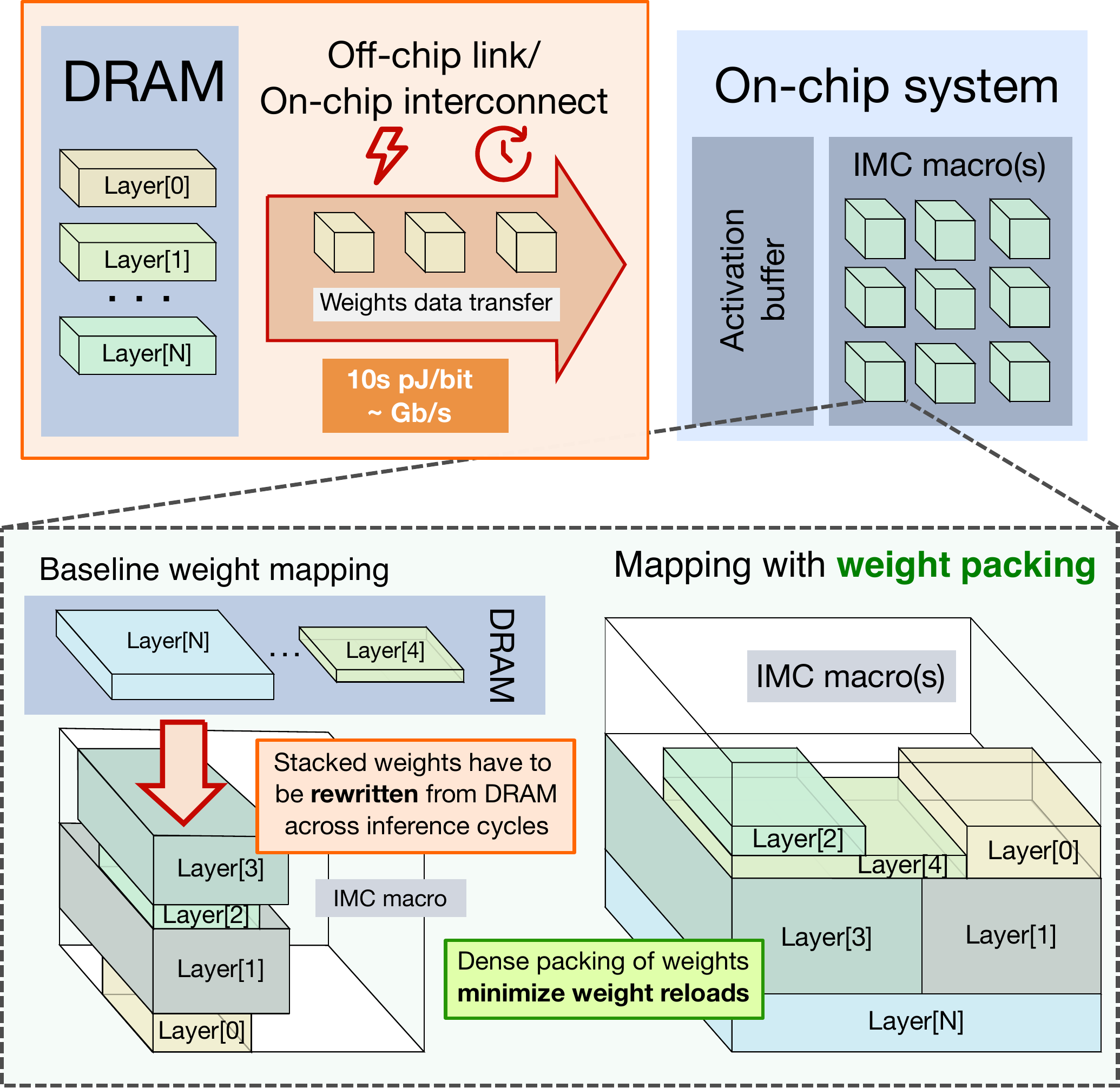}
    \caption{Weight reloading is a major energy and latency overhead in IMC computation for DNN workloads; the target of this work is to minimize its impact and maximize stationarity by packing efficiently the weights in the IMC array.}
    \label{fig:enter-label}
\end{figure}

For this, alternative hardware topologies have been investigated, and in-memory computing (IMC) has risen to prominence thanks to its peculiar features that make it a perfect candidate for hardware acceleration \cite{intro1, intro2, intro3}. IMC designs allow 1) massive matrix-vector multiplication parallelization thanks to the inherent structure of the memory array and 2) efficient data movement since the operands are efficiently reused spatially and temporally in the computation fabric. \cite{dse_1, dse_2}. The reduced impact of operand fetching from memory and the extremely low cost of the MAC operations enable up to 10$\times$ improvements in peak energy efficiency and latency. However, with real workloads IMC accelerators suffer from 1) underutilization of the available computational resources and 2) from weight reloading overheads \cite{dse_2}. Both of the mentioned points are heavily influenced by how the operands required for the MVM operations are stored in the memory array. By adopting suitable data layouts for the operands in memory and by densely packing the data in the IMC array the impact of the two degrading factors can be minimized. To face these issues it is required to act both on the hardware architecture and on the dataflow. From a hardware perspective, new IMC architectures include 1) multiple cells per multiplication unit to increase on-chip memory density (IMC designs present less memory density when compared to a conventional memory macro) and 2) multiple macros to increase dataflow flexibility and hence compute utilization.
Nevertheless, from a dataflow standpoint, a suitable mapping scheme for operands in novel IMC designs is still missing such that the available dense memory is optimally utilized -- minimizing thus data movement from and towards the IMC macros -- while at the same time not sacrificing throughput and energy efficiency of the computation. 

\begin{figure*}
    \centering
    \includegraphics[width=\linewidth]{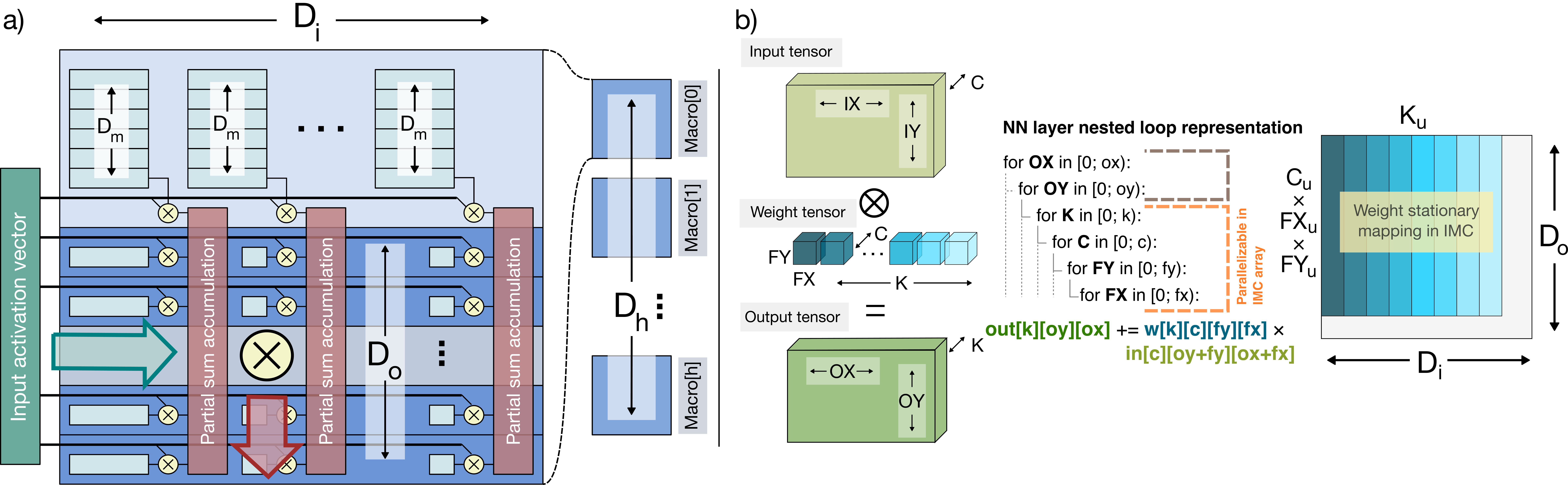}
    \caption{a) IMC template and its 4D design space. b) DNN operations and the weight stationary mapping of the weights in IMC}
    \label{fig:background}
\end{figure*}

The contributions of this work are the following: 
\begin{enumerate}
\item a mapping strategy to pack operands of different layers in the available memory in the IMC macros, while exploiting as much as possible the computational resources, towards minimizing the EDP of running edge AI workloads. 
\item A set of case studies on the MLPerf Tiny benchmark \cite{mlperftiny} on IMC architectures from literature to highlight when the proposed mapper can bring its most benefits, with potential $10-100\times$ improvements in EDP.
\end{enumerate}


\section{Background and overview}

\subsection{Dataflow concepts for IMC}
\label{sec:dataflow}
Deep neural network (DNN) workloads consist of a sequence of layers. The operations in the most common layers can be described as a combination of 6-nested for loops, which iterate over the indices of an input feature map tensor $I$, a weight tensor $W$ and that generate an output tensor $O$, as in Fig. \ref{fig:background}.b. The tensor operations can be decomposed in a sequence of matrix-vector multiplications (MVM) by tiling the suitable loops. These MVM operations offer a great opportunity for in-memory acceleration, as their dense 2D array structure aligns well with the array structure of the memory macros. In a single macro, the weight matrix is kept stationary in the array and the data layout in memory is configured such as to 1) maximize spatial reuse of the input activations, and 2) maximize accumulation of the partial sums. This is done by suitably spatially mapping \textit{irrelevant} loops for inputs and outputs respectively \cite{zigzag} in the array. The $K$ loop -- irrelevant for the inputs -- is unrolled across the input reuse dimension $D_i$, while the $C$, $FX$ and $FY$ loops -- irrelevant for the outputs -- are parallelized across the output reuse dimension $D_o$, as in Fig. \ref{fig:background}.b. Due to the hardware structure, within an IMC macro spatial mapping possibilities are limited to the weight stationary dataflow; however multiple IMC macros can be deployed in parallel. This allows for greater flexibility in the dataflow, at the cost of digital peripheral overhead: data movement and accumulation across the macros require an interconnection system and further glue logic. For this extra dimension we define a \textit{hybrid} dimension $D_h$. Across separate macros the input activation data can be programmatically multicast or unicast and outputs can be further accumulated or gathered based on the loops unrolled  across $D_h$. The space described by $D_i \times D_o \times D_h$ defines the maximum amount of spatial parallelism that can be achieved, and an IMC design achieves peak performance only when this space is maximally utilized -- within the weight stationary dataflow constraints. To increase memory density, multiple cells can be connected to a multiplier unit. The values stored in the cells can be time multiplexed across a further dimension $D_m$. This has been implemented in many designs recently \cite{imc1, imc2, imc9, imc13, imc14, imc15, imc16, imc17, imc18, imc19}
targeting edge applications. Fig. \ref{fig:background}.a contains a summary of the different dimensions considered in the design space.

\subsection{Motivation}

The development of the proposed mapping method for the weights is motivated by the need to reduce the impact of the two following contributions:
\begin{figure}[!t]
    \centering
    \includegraphics[width=0.6\linewidth]{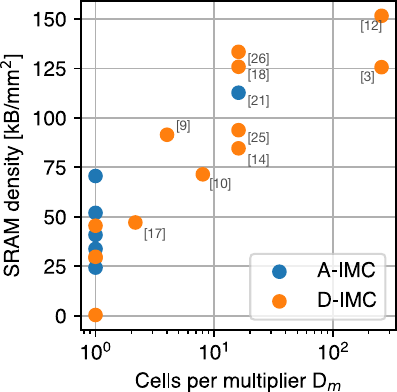}
    \caption{SRAM density increases proportionally with $D_m$; the contribution of multipliers and peripherals is amortized as we increase the number of cells per multiplier. This is adopted for both digital (D-IMC) and analog (A-IMC) designs.}
    \label{fig:dm_sram}
\end{figure}
\paragraph{Weight loading overhead} Weight loading affects both energy consumption and latency. Energy-wise, each loading requires fetching data from outside the IMC macro, reshuffle it so as to present it in the right alignment and load it in the memory array, with a large word parallelism. Latency-wise, weight loading and computation can not occur in parallel within one memory macro and this causes intrinsic stalls whenever the weight values have to be updated; beside this weights are often stored in off-chip memory (DRAM), characterized by insufficient bandwidth availabilities to avoid any stall and high access energies \cite{dram_cost}. To avoid this, weights should be kept as much as possible locally in the IMC array, avoiding repeated reloads across inference cycles; nevertheless to do so the amount of memory in the IMC arrays must be increased. An area efficient way to achieve this is to increase the $D_m$ dimension, by compactly stacking memory cells near each multiplier element; a cell density comparison survey is shown in Fig. \ref{fig:dm_sram}. We consider SRAM based IMC designs and compute the ratio of available memory compared to the IMC macro area; the designs are scaled according to the technology node of the chip, following \cite{sram_scaling}. In area constrained scenarios, it is shown that to expand on-chip memory, instead of increasing the number of macros $D_h$, increasing $D_m$ brings major improvements in SRAM density as the area impact of peripherals is amortized.

\paragraph{Underutilization of available computational parallelism}
The underutilization of the spatial parallelism when running the MVM operations hinders the performances and efficiencies of IMC designs.  As explained in Sec. \ref{sec:dataflow}, only a subset of loops can be effectively parallelized (namely $K, C, FX, FY$) in each layer across $D_i$, $D_o$. However loops can be parallelized across $D_h$ dimension still and contribute to latency and energy savings; even though this requires increased hardware complexity to handle accumulation or concatenation of the outputs across different macros, the benefits of exploiting low cost MAC operations far outweigh the periphery overheads.







In light of this, it is necessary to find a suitable mapping scheme for the weights in IMC architectures such that we can achieve 1) maximizing computational utilization and at the same time 2) mitigate weight writing overheads by maximizing IMC memory utilization.

\section{Weight Packing Algorithm}
To overcome the weight loading overheads without sacrificing computational parallelism, a weight packing algorithm is presented to tightly map the weights in the IMC macros. Given an IMC macro $D_i \times D_o \times D_h \times D_m$ and a workload, the objective is to minimize the total EDP = Energy$_{\text{total}}$ $\times$ Delay $_{\text{total}}$
    \begin{equation}
        \text{EDP}_{\text{total}} = \text{EDP}_{\text{MAC, Act. mem}} + \text{EDP}_{\text{Weight loading}}
    \end{equation}
when running inference of a network.

The MAC and the activation fetching/storing contributions are minimized by how largely the impact of peripheral elements is amortized \cite{dse_1}.  These include the circuitry for accumulation, for the propagation of activation data and the access to the activation memory. The amount of amortization is directly proportional to the amount of spatial parallelism that can be obtained in the IMC array in each computation cycle. In other words, the higher the spatial reuse that can be achieved for input and outputs in the IMC array across $D_i \times D_o \times D_h$ when doing MAC operations, the smaller will be the impact of activation memory data movement and IMC peripherals.


At the same time, fetching weights from an external memory -- particularly if off-chip -- heavily degrades performances and must be minimized by packing as many weights as possible in the available memory space inside the IMC macros
    

\begin{figure}[b]
    \centering
    \includegraphics[width=\linewidth]{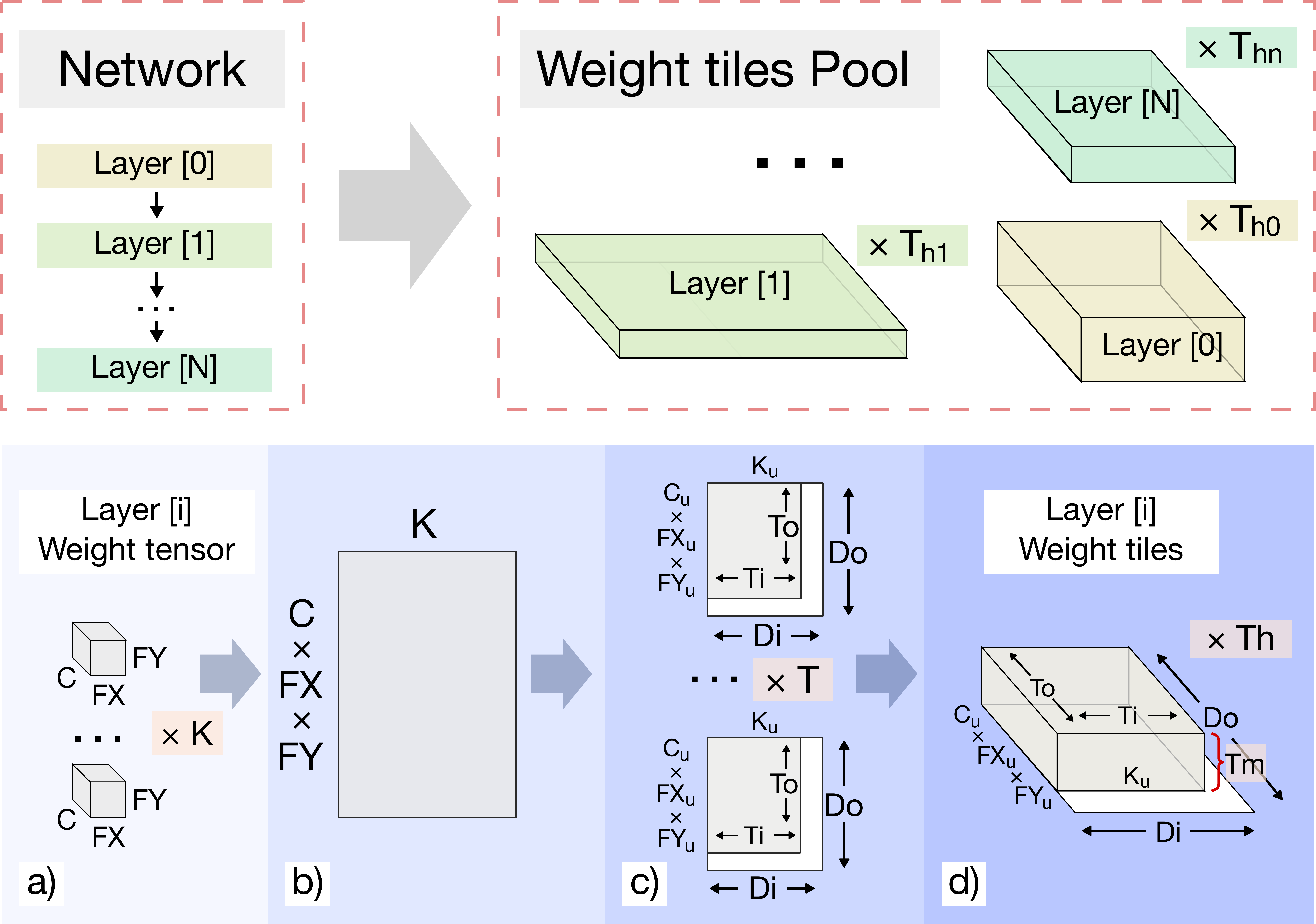}
    \caption{Weight tile pool generation steps}
    \label{fig:4}
\end{figure}

\begin{figure*}
    \centering
    \includegraphics[width=\linewidth]{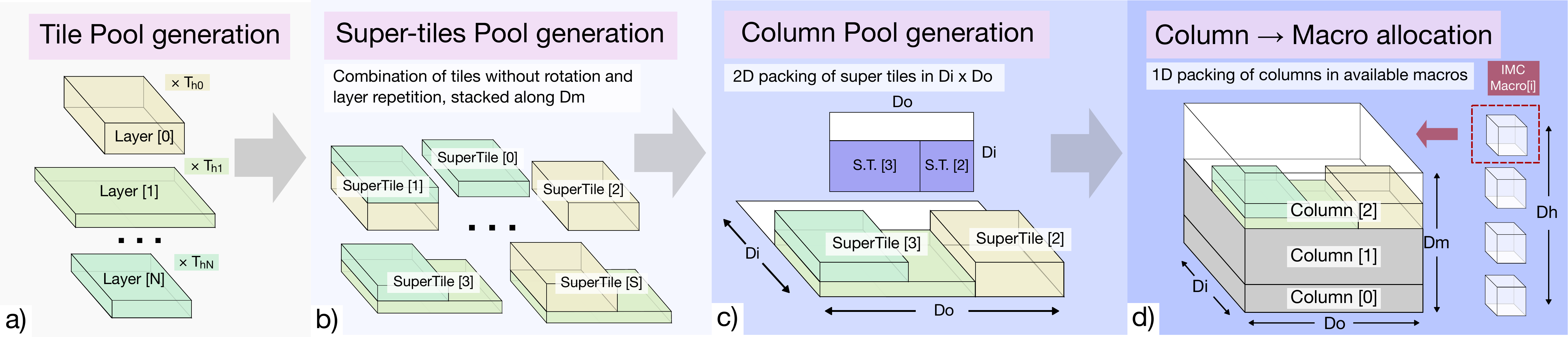}
    \caption{Steps successive to the weight tile pool generation: a) Supertile pool generation, b) column pool generation and finally 3) the allocation of columns to IMC macros.}
    \label{fig:5}
\end{figure*}

The two problems are intertwined:
\begin{itemize}
    \item Maximizing memory utilization translates in occupying the available memory space across $D_i \times D_o \times D_h \times D_m$ such as to minimize weight reloading.
    \item Maximizing compute utilization requires maximizing the utilization across $D_i \times D_o \times D_h$ during each computation cycle when running the layers in a network. This has to occur under the stringent constraints of the weight stationary dataflow (Fig. \ref{fig:background}.b), which limits the allocation possibilities.
\end{itemize}


The tiling and packing algorithm presented is a 3-dimensional bin-packing problem, a multi-dimensional version of the bin-packing problem well known for being NP-hard \cite{bin_packing}.
By means of heuristics and by applying constraints specific to this use case, we are able to solve it by splitting it into a 2D + 1D packing problem and achieve tight packing of the weights in the memory fabric. The overview of the steps required to run the packing algorithm are described in Fig. \ref{fig:4} and Fig. \ref{fig:5}. 

Firstly a pool of weight tiles is identified based on the IMC dimensions; to improve the packing density the pool is extended to include combinations of original tiles (\textit{supertiles}), as in \cite{3dbpp}. The elements in the pool thus have to be allocated in the IMC macros: this is done by means of a \textit{column} generation step, followed by the allocation of the columns in different macros. The former step consists in identifying dense allocation of supertiles in the $D_i \times D_o$ space (\textit{columns}), while the latter step places the columns in different IMC macros across the $D_h \times D_m$ dimensions. The details of each step are explained in the following sections.

\subsection{Tile generation}
\label{sec:tile_generation}
The first step requires the definition of an initial set of weight tiles: for each layer a set of uniform tiles is found that fits in $D_i$, $D_o$, $D_h$, $D_m$ . The dimensions of the tiles have analogous names: for each layer there are $T_{h}$ tiles of size $T_{i} \times T_o \times T_m$, referring respectively to the input reuse, output reuse and time multiplexing tile dimensions (as per Section 2). The initial dimension of the tiles ($T_i \times T_o$) are such that $T_i$ (and $T_o$) are the combination of loop prime factors (LPFs) \cite{zigzag} for $K$ ($C$, $FX$, $FY$ respectively) that maximize utilization across $D_i$, ($D_o$ respectively), as in Figure 4 (step c).   

The LPFs that are not unrolled across $T_{i}$, $T_o$ have to be distributed across $D_h$ and $D_m$. To maximize spatial parallelism (and thus compute utilization), the LPF combination that maximizes utilization across $T_h$ is firstly identified: the \textit{input relevant} LPFs are prioritized (those related to the $C$, $FX$, $FY$ dimensions \cite{zigzag}) -- as they contribute to higher spatial reuse for the partial sums, while the output relevant LPFs are selected when the input relevant ones are exhausted. (step c) 
Those LPFs that are not spatially unrolled across $D_h$ are left to be temporally multiplexed across $T_m$ (step d).



\subsection{SuperTile Generation}
Supertiles are a combination of stacked tiles without rotation in the $D_m$ dimension, similar to the concept of superitems described in \cite{3dbpp}. As shown in Fig. \ref{fig:5}.b, the same tile from the original tile pool can be found in different supertiles. However the tile stackings are not an exhaustive set of all possible combinations but constrained by:

\begin{enumerate}
    \item The stack in $D_m$ dimension contains at most one tile per layer: this is done so as to maximize the spatial parallelism of tiles of the same network layer across $D_i \times D_o \times D_h$ and avoid losing spatial parallelism
    \item The cumulative height of the stacked tiles ($\sum T_m$) does not exceed the largest $T_m$ in the original tile pool; this is a lossless heuristic adopted to speed up the search. 
\end{enumerate}
Each supertile generated is now thus characterized also by a new set of dimensions, $ST_i$, $ST_o$, $ST_m$. $ST_i$ and $ST_o$ are derived based on the largest tiles and $ST_m$ as the sum of all $T_m$ belonging to the tile stack.  


\subsection{Column generation}
Once the pool of supertiles is defined, it is necessary to find their densest allocation across the available IMC macros (maximizing memory utilization), without sacrificing spatial parallelism of each network layer (maximizing compute utilization). The compute maximization objective implies that tiles that belong to the same layer must be spatially parallelized across IMC macros ($D_h$), and should not be stored in the same macro.

The allocation is done iteratively: a subset of supertiles is selected with tiles belonging to different network layers. The selected supertiles are then packed into columns: if the 2D packing of supertiles across the $D_i \times D_o$ dimensions succeeds, the density of the found supertile allocation is computed and compared to the densest allocation found up until that point. The density of a column allocation is computed as the ratio between the sum of all the tiles volumes and ($D_i \times D_o \times ST_{m, \text{max}}$), with $ST_{m, \text{max}}$ being the $D_m$ occupation of the largest supertile in the allocation.

Once all the possible allocations are evaluated and the densest combination is identified, a column is generated and the supertiles that compose the newly generated column are removed from the pool. The process is then repeated until the pool is empty.
 
By doing so densely packed columns are sequentially generated, composing a pool of columns to be allocated in the IMC macros.

\subsection{Column allocation to macros}
The set of identified columns has to be placed across the $D_h \times D_m$ space, within and across the IMC macros. The allocation is solved as a constrained 1-D bin packing problem, with the constraint being the requirement to pack at most one tile of a layer per IMC macro, thus distributing tiles of the same layer across $D_h$, so as to maximize compute utilization.

\begin{figure}
    \centering
    \includegraphics[width=\linewidth]{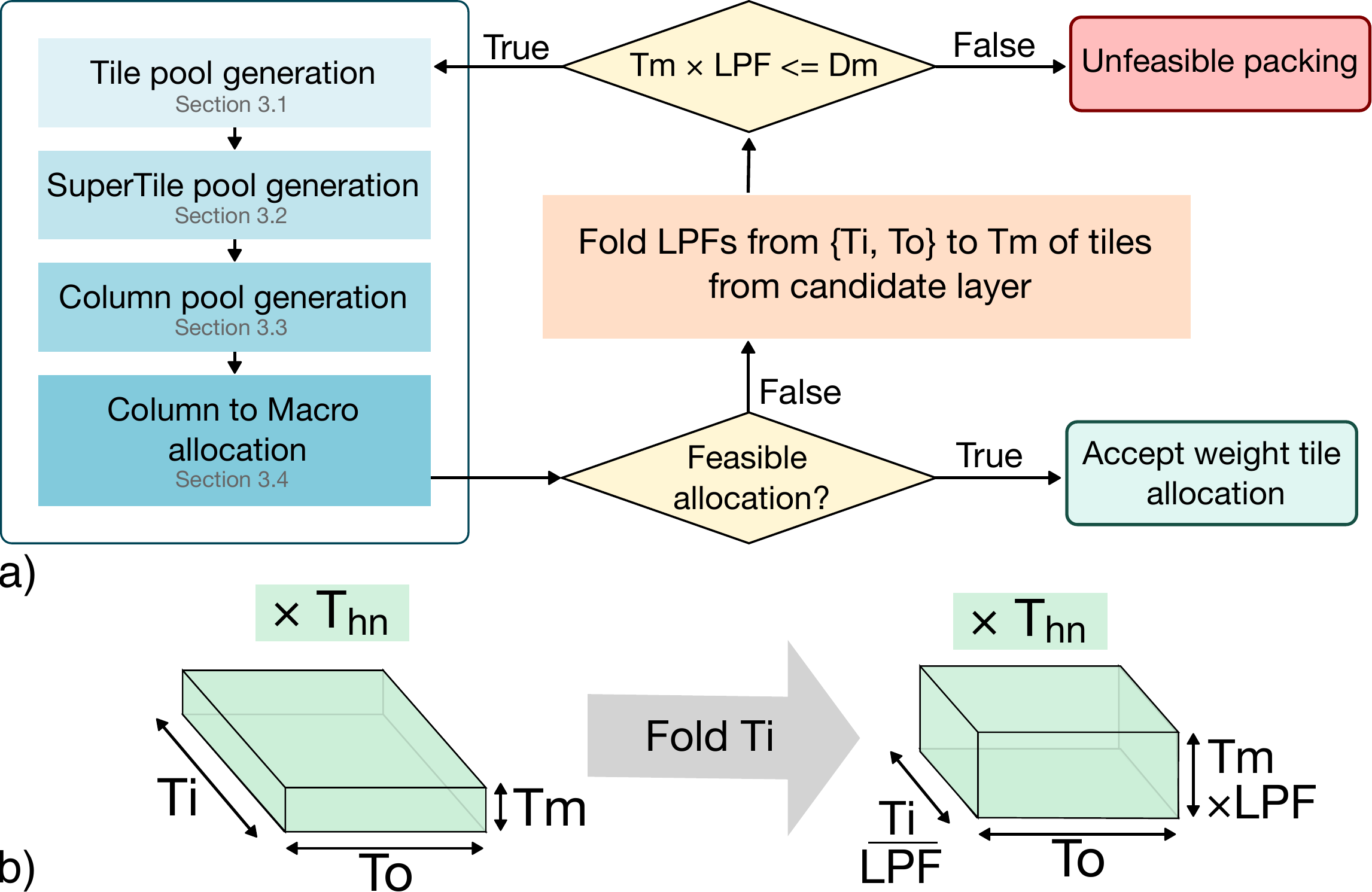}
    \caption{a) Weight packing flow summary and b) an example of a folding step, where the $T_i$ dimension of the original tile is divided by the LPF, which is in turn folded in the $T_m$ dimension.}
    \label{fig:packing_flow_summary}
\end{figure}

In the case is not possible to allocate the columns in the available macros, the \textit{folding} strategy is considered. Tile folding in $D_m$ consists in firstly identifying a candidate layer and subsequently selecting a LPF from $T_i$ or $T_o$ and transforming it from a spatial unroll to a temporal loop in $D_m$, as in Fig. \ref{fig:packing_flow_summary}.b. This operation effectively reduces the footprint of the tiles across $D_i \times D_o \times D_h$, while increasing the $T_m$: by doing so the tile allocation possibilities increase, since the $D_m$ dimension is less constrained than $D_i$, $D_o$, $D_h$. The layer candidate for tile folding is chosen as the one with the \textit{lowest latency} with the given tiling configuration.
This heuristic is selected with the premise that layers with lower computation time also have larger weight tensors, and thus folding would greatly reduce footprint of the tile in $D_i \times D_o$ while causing the smallest increase in latency. It is effectively observed in most network architectures that as we go deeper in the network the weight tensors are expanded while the $OX$ and $OY$ dimensions of activation tensors -- the only ones that can not be parallelized efficiently in IMC -- become smaller. The tile dimensions of the layer with the lowest inference latency are folded from the $T_i, T_o$ dimension to $T_m$ according to the smallest LPF available. If the folded tile $T_m$ exceeds available $D_m$, the next lowest latency layer is chosen for folding. If no layer can be found that can be effectively folded, the packing is deemed unfeasible. Folding of $K_u$ loops is prioritized over output irrelevant spatially unrolled loops as the former  cause temporal stationarity for the inputs, avoiding multiple re-fetches of the input activations from the local buffer.  



\section{Case studies}
The mapper is integrated in ZigZag-IMC \cite{zigzag-imc} to carry out a set of case studies to highlight the benefits of the mapper under different types of workloads and the trade-offs with conventional mapping schemes. 
Two IMC architectures from literature are considered for the case studies: a digital IMC (D-IMC) \cite{dimc_exp}
and an analog IMC (A-IMC) \cite{aimc_exp}. The unit costs required for energy and area estimation are extracted from their reported peak performances and summarized in Tables \ref{tab:parameters}. The values are then plugged-in in the ZigZag-IMC cost model to extract efficiency estimates. Furthermore, to include system-level contributions to the estimates, a 256kB  on-chip buffer is considered for storing intermediate activation data and an external LPDDR4 memory to fetch weights. The unit costs and bandwidths of the memories are summarized in Table \ref{tab:parameters}. 

In a first study, we compare the proposed mapping method for the weights to two conventional mapping methods and show fundamental benefits and trade-offs of the weight packing method. 

In a second study, we analyze area vs. EDP trade-offs by sweeping for a set of D$_h$, D$_m$ configurations on target edge workloads.

\begin{table}[t]
  \caption{Baseline hardware parameters}

  \label{tab:parameters}
  \begin{tabular}{c|c|c|c}

  \multicolumn{4}{c}{\textbf{22nm D-IMC design \cite{dimc_exp}}}\\
    \toprule

    $D_o \times D_i$ & 256 $\times$ 16 & Macro Area [mm$^2$]  & 0.202 \\
    $D_h \times D_m$ & 1 $\times$ 1  & Cell area [$\mu$m$^2$]  & 0.379\\
    Operand prec. & 4bW/4bI & Periph. area [$\mu$m$^2$] & 44290 \\
    Operating point & 0.9V@200MHz & ND2 cap. [fF] & 0.3\\
  \bottomrule

  \multicolumn{4}{c}{\textbf{28nm A-IMC design \cite{aimc_exp}}}\\
    \toprule

    $D_o \times D_i$ & 256 $\times$ 16 & Macro Area [mm$^2$]  & 0.035 \\
    $D_h \times D_m$ & 1 $\times$ 1  & 10T Cell area [$\mu$m$^2$]  & 1.2 \\
    Operand prec. & 4bW/4bI & Periph. area [$\mu$m$^2$] & 15400 \\
    Operating point & 0.9V@200MHz & ADC conv. [fJ/conv] & 190\\
  \bottomrule
\multicolumn{4}{c}{\textbf{Memory instances}}\\
      \toprule
   & LPDDR4 \cite{dram_cost} &  \multicolumn{2}{c}{256kB  SRAM buffer \cite{cacti}} \\

    \midrule
    
    Energy R/W [pJ/bit] & 4 & \multicolumn{2}{c}{0.009} \\
    Bandwidth [Gb/s] & 12.8 & \multicolumn{2}{c}{/} \\
  \bottomrule
\end{tabular}
\end{table}

 \begin{figure}[b]
    \centering
    \includegraphics[width=0.8\linewidth]{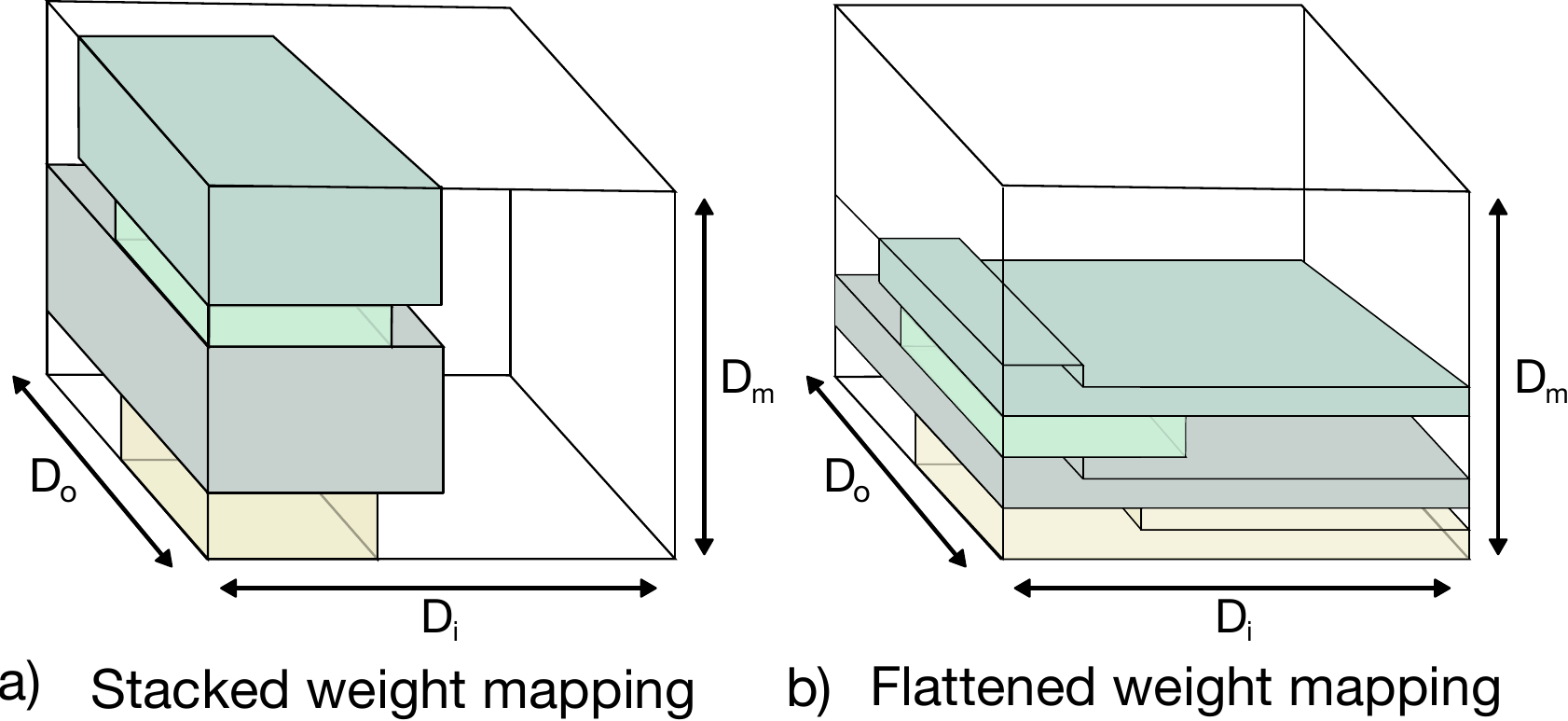}
    \caption{Baseline mapping methods from literature}
    \label{fig:mapping_comparison_fig}
\end{figure}
 
\subsection{Weight mapping methods comparison}

We distinguish two baseline types of weight mapping methods: \textit{stacked} and \textit{flattened}. In the \textit{stacked} method (as in \cite{imc_design_6}) the weight tile dimensions are defined as a combination of LPFs and are uniform, meaning that all tiles have the same shape -- effectively obtaining the pool of tile described in Sec. \ref{sec:tile_generation}. However, no packing is applied, and the tiles are stacked vertically on top of each other, within $D_m$  (Fig. \ref{fig:mapping_comparison_fig}.a). The second method -- \textit{flattened} -- assumes that the  weight tensor is spread across $D_i$ and $D_o$ as much as possible, eventually folded across $D_m$, even in non-uniform tiles, as in Fig. \ref{fig:mapping_comparison_fig}.b. 

The described mapping methods are compared with the proposed method when applied on the hardware baseline described ($D_o \times D_i = 256 \times 16$) and with one single macro ($D_h = 1$). The workloads considered are from the MLPerf Tiny Benchmark \cite{mlperftiny} as they feature a comprehensive variety of layer shapes and dimensions and showcase the flexibility of the presented mapping method. 

\begin{figure}[t]
    \centering
    \includegraphics[width=0.8\linewidth]{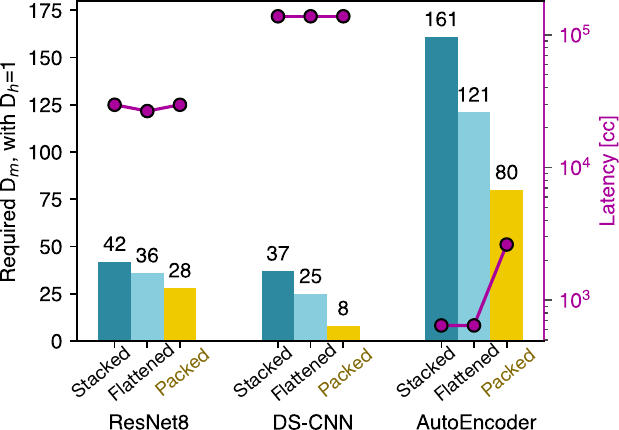}
    \caption{EDP comparison between the baseline mappings of Fig. \ref{fig:mapping_comparison_fig} and the proposed packed mapping of the weights.}
    \label{fig:mapping_comparison}
\end{figure}

The results are summarized in Fig. \ref{fig:mapping_comparison}. We show that in all cases the proposed mapping outperforms the previous  methods when considering the minimum required $D_m$ for mapping the whole network -- thus area efficiency and memory utilization -- as the main metric for evaluation. The packed method is particularly effective when considering networks characterized by small weight tensors with respect to the available $D_i \times D_o$, such as in DS-CNN. 

However the packing and the folding operation required in some cases have a direct impact on the latency required for the computation: by folding we effectively translate a spatially parallelized loop to a temporal loop, increasing proportionally the number of clock cycles required for the computation. This can be observed in the case of the AutoEncoder network: tight packing of the weights can be achieved in the available memory space but at the cost of folding multiple times the weights in $D_m$.

\subsection{Impact of weight loading and $D_h$}

\begin{figure}
    \centering
    \includegraphics[width=\linewidth]{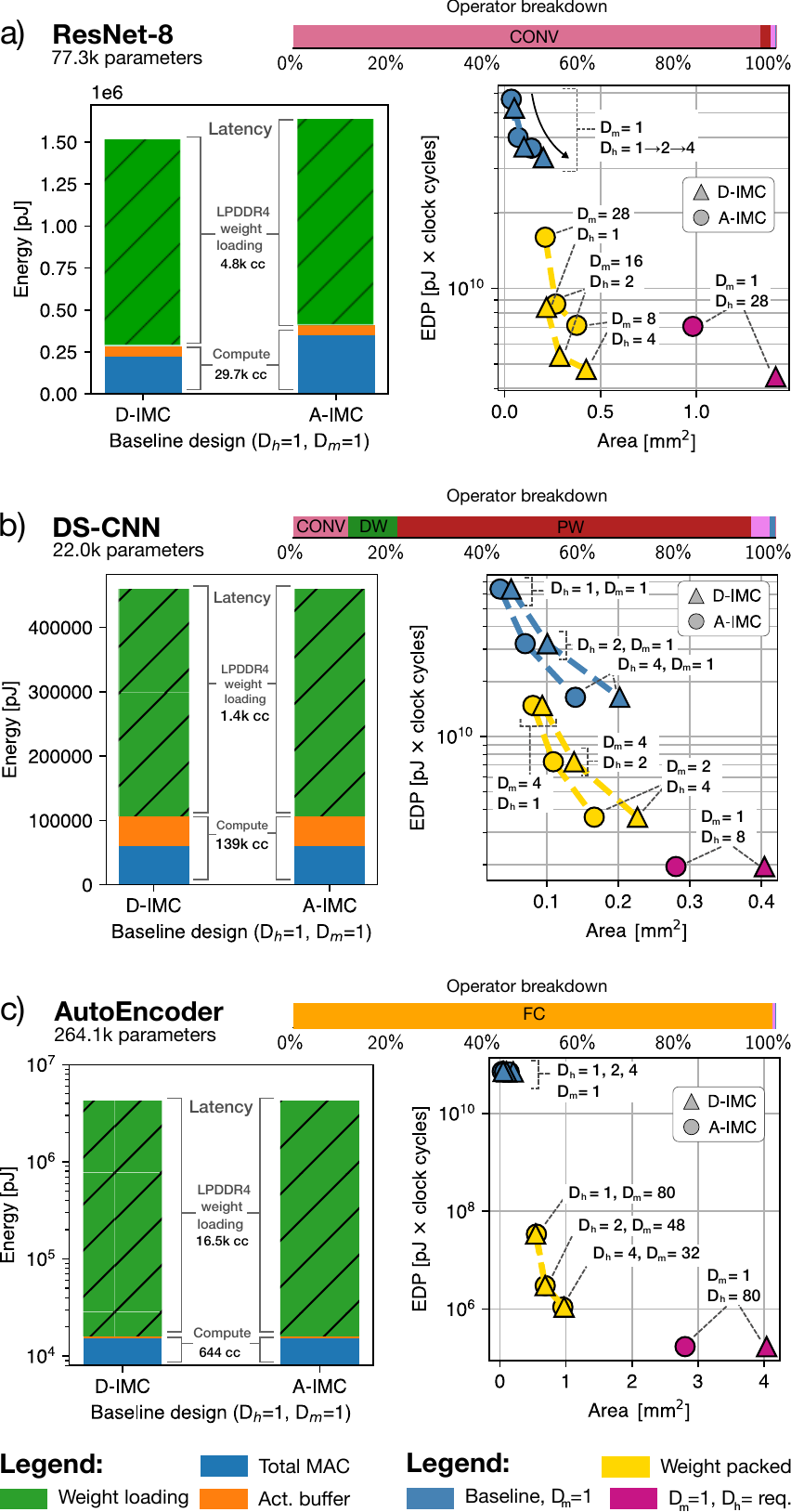}
    \caption{EDP vs area trade-offs between the stacked mapping and the proposed packed mapping of the weights.}
    \label{fig:second_exp}
\end{figure}

We can thus use the implemented mapping algorithm and integrate it in ZigZag-IMC \cite{zigzag} to obtain system level estimates.
To evaluate the impact of the weight loading from DRAM and area and EDP trade-offs of different design points we consider the D-IMC \cite{dimc_exp} and A-IMC \cite{aimc_exp} designs selected as baselines and we cover different $D_m$, $D_h$ combinations. We consider as workloads the ones of the previous study. The results are reported in Fig. \ref{fig:second_exp}, with comprehensive details of each point. The starting point for the estimates is the architecture from the publication, thus with $D_h, D_m = 1$. The energy consumption and the latency of the three major contributions are reported on the left hand side plots of Fig. \ref{fig:second_exp}. As expected, in all cases the loading of the weights from DRAM is a heavily detrimental contribution to the performances of the accelerator. The contribution of the on-chip activation buffer is not dominant compared to the MAC compute because of the high spatial data reuse of the operands in the IMC array and because of the efficiency of the MAC operations in the array. By increasing the number of macros ($D_h = 1,2,4$, $D_m=1$) the parallelization possibilities increase proportionally, but the EDP benefits still dwarf compared to the weight loading overhead. This is observed on the blue trace of the right hand side figure. If however we increase the number of cells per multiplier $D_m$ and we sweep for the same set of $D_h$ settings, we observe that, at the cost of a fraction of mm$^2$ of extra area, we are able to compactly pack the weights and avoid any cost associate with reloading from DRAM. This scenario corresponds to the yellow trace of the left hand side plots and is where the proposed mapping method is applied. If on the other hand we fix $D_m=1$ and we increase the number of macros $D_h$ to the required amount such as to be able to 2D pack the weights, we observe that, by paying the price of $>$1-2$\times$ increased IMC area compared to the packed solution, there is intrinsically no folding of the weight tiles and thus some marginal consequent EDP benefits. (Purple trace in Fig. \ref{fig:second_exp}). Considering that for larger networks with millions of parameters the increase in area required would make it unfeasible to parallelize across $D_h$, the weight packing solution provides a viable solution to mantain all weights on chip, at the expense of a minor loss in EDP.

\section{Conclusion}

This paper presents a novel method to tightly pack the weights of NN workloads in IMC array across a 4-dimensional space defined by $D_i \times D_o \times D_h \times D_m$. We show that the proposed method can erase the impact of weight reloading overheads with a small area overhead. Through a series of case studies it is shown that the mapping method proposed outperforms baseline methods from literature and provides significant EDP benefits compared to conventional mapping, with up to $\sim100\times $ EDP improvements for weight dominant workloads.

\bibliographystyle{ACM-Reference-Format}
\bibliography{sample-base}

\end{document}